
\font\fontbig=cmr10 scaled \magstep2

\def\L{{\cal L}}

\mathchardef\lag="724C
\magnification=1200
\hsize=16.0 true cm
\baselineskip 15pt

\def\pl{{\sl Phys.\ Lett.\ }}
\def\np{{\sl Nucl.\ Phys.\ }}
\def\pr{{\sl Phys.\ Rev.\ }}
\def\prl{{\sl Phys.\ Rev.\ Lett.\ }}

\rightline{BIHEP-TH-95-33}
\rightline{December 1995}
${}^{}$
\vskip 4pc
\centerline{\bf {\fontbig TOWARDS YUKAWA UNIFICATION IN A}}
\centerline{\bf {\fontbig STRONGLY YUKAWA COUPLING MODEL}}
\vskip 4pc
\centerline{\bf Keyan Yang }
\vskip 1pc
{\sl
\centerline{\bf Institute of High Energy Physics, Academia Sinica}
\centerline{\bf Beijing~100039, China}
}
\vskip 4pc
\centerline{\bf ABSTRACT }
\vskip 2pc

For strong enough Yukawa coupling the electroweak standard model
fermion finds it energetically advantageous to transform itself
into a bound state in the hedgehog background of Higgs field
in semiclassical approximation.
By considering the bound states give the masses for lepton and quark,
we found that all the strongly Yukawa coupling constants
tend to an unitary constant.
On the other hand, the masses of $W$ and $Z$ produced by the hedgehog
Higgs field keep almost no change.
\vskip 2pc
\par
PACS number(s): 12.50.Ch, 12.15.Ff, 11.10.St

\vfil\eject

The standard model of the electroweak interaction[1] has received a great
deal of phenomenological support. The gauge boson and fermion structure of
the model has been confirmed to a high degree of accuracy, while, there is
very little phenomenological support for the Higgs sector of the theory.
The most difficult part to understant of this theory is the Yukawa sector.
It contains a large number of free parameters that must be adjusted by hand
to obtain a realistic spectrum of particles and mixings.
Recently, it is reported that the top quark mass is much bigger than the
masses of other lepton and quark[2].
There is a large mass hierarchy among three generation fermions.
These features of the standard electroweak theory may be understanted by use
of some of the successful ideas of this theory, or by entirely new ideas.
\par
The basic dynamic degrees of freedom in the electroweak
theory are the two-component Weyl fields with definite helicities.
The gauge interactions conserve helicity and do not mediate between
the system of the lefthanded and the one of the righthanded fermions.
The only bridge between those two is provided by the hypothetical scalar
Higgs boson which couples to the various fermions with a strength
proportional to their masses.
The lepton and quark in standard model are massless before
the spontaneous symmetry breaking (SSB) and their masses are produced by
the Yukawa coupling after the SSB.
The standard picture in the model with SSB involves a vacuum expectation
value (VEV) for the Higgs fields which are constant over the whole space.
The fermion mass values are given by introducing Yukawa coupling constants,
and the masses are proportional to the Yukawa coupling constants.
On the other hand, it is expected that the presence of fermion with
strong coupling to the Higgs field can modify the standard picture.
As the Yukawa coupling increases the Higgs field around a fermion tends to
go over from a uniform spatial configuration to a hedgehog configuration[3].
The uniform configuration leads to a fermion mass which increases linearly
with the coupling constant while the hedgehog configuration gives a
decreasing mass.
\par
The existence of two possible configurations means that one can get
the same mass value in both weakly Yukawa coupling phase and strongly
Yukawa coupling phase.
For describing the mass spectrum of lepton and quark in the weakly coupling
phase, we must introduce nine hierarchical Yukawa coupling constants
which vary about from $10^{-6}$ to 1.0.
In this letter, alteratively, we shall consider the consequences of
strongly Yukawa coupling phase to take the place of the standard picture
for fermion sector in Weinberg-Salam model.
We found that in the strong coupling phase all the Yukawa coupling
constants tend to an unitary constant and the hierarchy of Yukawa coupling
constants is absent.
Furthermore, this scheme gives a hint to unify the all Yukawa coupling
constants.
\par
The strongly coupled Higgs fermion sector of the standard model has been
investigated in Ref. 3. In those works they exploit the analogy between the
the strongly Yukawa coupling theory and the chiral linear $\sigma$-model.
They follow the papers of Ref. 4 which work in the context of strong
interaction.
In this letter, we will focus on some features of this scheme for
the strong Yukawa coupled standard model from a different point of view.
We change the strength of Yukawa coupling in the standard electroweak theory
and show that the fundamental ground state of the fermion-Higgs system is not
the trivial ground state which is given by the uniform spatial Higgs field
and is the nontrivial ground state with hedgehog configuration of Higgs field.
The nontrivial ground states give the masses to lepton and quark.
\par
Throughout this letter we work in the classical approximation, i.e.
we neglect the radiative corrections due to the boson loops and the
contribution of the Dirac sea to the energy of the system.
The question of whether these effects are important is beyond the scope of
the present letter.
On the other hand, the large Yukawa constants in the standard model cannot
be included without violating vacuum stability[5]. However, the problem may
be overcome by introducing new physics in the TeV region, for example,
supersymmetric theories.
Therefore, we think that our results for strong coupling model can be
qualitatively correct.


\par
We shall first make a brief review on the electroweak standard model.
The model contains left-handed and right-handed Weyl fermion fields and
$SU(2)_L\times U(1)$ gauge fields $A_\mu^a$ and $B_\mu$ and a complex Higgs
doublet $\Phi$. The Lagrangian of the model is given by $\L=\L_b+\L_f$.
The bosonic Lagrangian is
$$
     \L_b =-{1 \over 4}F^a_{\mu\nu}F^{\mu\nu,a}
          -{1 \over 4}G_{\mu\nu}G^{\mu\nu}+
          (D_\mu \Phi)^\dagger (D^\mu \Phi) - \lambda
          (\Phi^\dagger \Phi - {v^2 \over 2} )^2
          \eqno (1)
$$
where $F^a_{\mu\nu}$ and $G_{\mu\nu}$ are the $SU(2)_L$ and $U(1)$ fields
strength tensor, respectively.
The fermionic Lagrangian in the chiral representation reads,
$$
  \eqalign{  \L_f = &{\bar \psi}_L i\gamma^\mu D_\mu \psi_L+{\bar u}_R
                     i\gamma^\mu D_\mu u_R
                   + {\bar d}_R i\gamma^\mu D_\mu d_R       \qquad  \cr
                 &-f_U( {\bar \psi}_L {\tilde \Phi} u_R
                       +{\bar u}_R {\tilde \Phi}^\dagger \psi_L)
                -f_D( {\bar \psi}_L \Phi d_R+{\bar d}_R \Phi^\dagger \psi_L)
}     \eqno (2)
$$
where $\psi_L$ denotes the left-handed doublet $(u_L, d_L)$, $u_R$ and $d_R$
are the right-handed singlets,
and $\tilde\Phi=i\tau_2\Phi^{*}$. Here, we use ($u$, $d$) to represent
(up, down) lepton pair and quark pair. In order to discuss the strongly
Yukawa coupling model in the follow, we introduced right-handed neutrino
to make the neutrino small Dirac mass.
\par
The $SU(2)_L$ gauge symmetry is spontaneously broken due to the non-vanishing
VEV $v$ of the Higgs field,
$$
      \langle \Phi \rangle =-{v \over \sqrt2}{0\choose 1}   \eqno (3)
$$
leading to the gauge boson $W$ and $Z$ masses and Higgs mass:
$$
     M_W={1 \over 2}g_2v, \quad  M_Z={1 \over 2}(g_1^2+g_2^2)^{1 \over 2}v,
     \quad M_H=v\sqrt{2\lambda}, \eqno (4)
$$
where $g_1$ and $g_2$ are $U(1)$ and $SU(2)$ gauge coupling constant,
respectively, and we take $v$=246.0GeV.
When the Higgs field develops the VEV $v$, we see that
the Higgs field configuration takes the form of spatially uniform field.
The fermions then acquire masses through their Yukawa couplings $f_U$ and
$f_D$:
$$
      m_U^{}=f_U{v \over \sqrt2}, \qquad m_D^{}=f_D{v \over \sqrt2}.
            \eqno (5)
$$
The masses of three generations lepton and quark in the Weinberg-Salam model
are described by the following Yukawa coupling constants:
$$
\eqalign{
      &f_{\nu_e}<2.93\times 10^{-11}, \quad f_e=2.94\times 10^{-6},
      \quad f_u=2.87\times 10^{-5}, \quad f_d=5.75\times 10^{-5},  \cr
      &f_{\nu_\mu}<1.55\times 10^{-6}, \quad f_\mu=6.07\times 10^{-4},
      \quad f_c=7.47\times 10^{-3}, \quad f_s=1.15 \times 10^{-3},  \cr
      &f_{\nu_\tau}<1.78\times 10^{-4}, \quad f_\tau=1.02 \times 10^{-2},
      \quad f_t=1.01\times 10^{+0}, \quad f_b=2.47\times 10^{-2},
}   \eqno (6)
$$
where we have taken the approximate values for $m_u=5$MeV, $m_d=10$MeV,
$m_s=200$MeV, $m_c=1.3$GeV, $m_b=4.3$GeV and $m_t=176$GeV.
We see that the Yukawa coupling constants are unorganized in a wide range.
A large hierarchy appeares in the Yukawa coupling constants.
In order to decrease the number of Yukawa coupling parameteres,
it is much nessceary to unificate the Yukawa coupling constants.


\par
To this end, instead of the weakly Yukawa coupling in the Weinberg-Salam
model, let us consider the case of the strongly Yukawa coupling.
We would like to compare its predition with a ground state energy estimated
in a semi-classical manner.
Due to the Higgs field gives the mass of the fermions, we can neglect
the gauge degree of freedom with $A_\mu^a=B_\mu=0$ in Weinberg-Salam model.
By a simple redefinition of Higgs fields
$$
   \Phi={1 \over \sqrt2}{\pi_2+i\pi_1  \choose \sigma-i\pi_0}   \eqno(7)
$$
and fermion fields $\psi=\psi_L+\psi_R$ where $\psi_R$ stands for $u_R$ and
$d_R$, then, the Lagrangian $\L$ can be written as $\L=\L_\sigma+\L'$ with[3]
$$
   \L_\sigma ={1 \over 2}(\partial_\mu\sigma)^2
                 +{1 \over 2}(\partial_\mu\displaystyle{\bf \pi})^2
                 +\bar \psi\gamma^\mu\partial_\mu \psi
                 -g\bar \psi(\sigma+i\gamma_5
                 \displaystyle{\bf \pi}\cdot\displaystyle{\bf \tau})\psi
                 -U(\sigma, \displaystyle{\bf \pi})    \eqno (8)
$$
where
$$
     U(\sigma, \displaystyle{\bf \pi})=
            {\lambda \over 2}\big [\sigma^2
            +\displaystyle{\bf \pi}^2 -v^2\big ]^2,
               \eqno(9)
$$
and
$$
   \L' = {f_U-f_D \over 2\sqrt 2}[i\pi_0\bar \psi(1-\gamma_5)\psi
           +\bar \psi\tau_3(\sigma
           +i\displaystyle{\bf \pi}\cdot\displaystyle{\bf \tau})\psi]
            \eqno(10)
$$
where $g=(f_U+f_D)/2\sqrt2$ and $\L_\sigma$ is the $\sigma$-model
Lagrangian. In the following we will see that in the strong coupling model
all the Yukawa coupling constants tend to an unitary constant and
$(f_U-f_D)$ tends to zero. So, in what follows we shall, then,
as a simplification, neglect $\L'$ leaving us with $\L=\L_\sigma$ and
set $f_U=f_D=f$.


\par
There is no guarantee that the classical solution (3) gives the lowest
possible ground state energy in the fermion sector. In the semi-classical
approximation we include only valence fermion and we treat the
$(\sigma, \displaystyle{\bf \pi})$ fields as classical.
We consider the hedgehog solution for which the chiral fields have the
form in terms of the fields:
$$
      \sigma(x)=\sigma (r), \quad
      \displaystyle{\bf {\pi}}(x)=\pi(r)\hat{\bf r}      \eqno (11)
$$
with $\hat{\bf r}={\bf r}/r$.
The pion field has then a dipole shape whose space orientation is coupled
to the isospin orientation.
\par
With chiral fields of the form (11), the Dirac equation of fermion
admits an s-state solution of the form
$$
    \psi(x)={1 \over \sqrt{4\pi}}
            {G(r) \choose i({\bf \sigma}\cdot\hat{\bf r})F(r)}\chi_h,
               \eqno (12)
$$
where $\chi_h$ is a state in which the spin and isospin of the fermion
couple to zero as $({\bf \sigma}+{\bf \tau})\chi_h=0$.
Taking the ansatz of (11) and (12), minimization of (8) produces the
following coupled non-linear equations:
$$
\eqalign{
   &{1 \over r}{d^2 \over dr^2}[r\sigma(r)]=
        {\partial U \over \partial \sigma}
        -{g^2 \over 4\pi}[G^2(r)-F^2(r)],    \cr
   &{1 \over r}({d^2 \over dr^2}-{2 \over r^2})[r\pi(r)]=
        {\partial U \over \partial h}
        +{g^2 \over 2\pi}G(r)F(r), \cr
   &{dF(r) \over dr}+[{2 \over r}-g\pi(r)]F(r)
                    +[\omega+g\sigma(r)]G(r)=0,  \cr
   &{dG(r) \over dr}+g\pi(r)G(r)+[-\omega+g\sigma(r)]F(r)=0.
}  \eqno (13)
$$
The quantity $\omega$ is the eigenvalue of the spinor $\psi(x)$
and the radial fermion wave functions are normalized to
$\int r^2dr[F^2(r)+G^2(r)]=1$.
\par
These equations are supplemented by the following boundary condition:
$$
\eqalign{
        &\sigma(r)\to c_1, \quad \pi(r)\to 0,
         \quad G(r)\to c_2, \quad F(r)\to 0,
         \quad {\rm as} \quad r\to 0;          \cr
        &\sigma(r)\to -v, \quad \pi(r)\to 0,
         \quad G(r)\to 0, \quad F(r)\to 0,
         \quad {\rm as} \quad r\to \infty
}  \eqno (14)
$$
where $c_1$ and $c_2$ are arbitrary constants.
These express the fact that the physical vacuum is recovered at
infinity. In this 'physical' vacuum the fermions are free Dirac particles
of mass $gv$, and chirality is spontaneously broken.
\par
The hedgehog solution is a self-consistent solution in the sense that,
when fermions occupy an orbital of the form (12), the equations in (13)
for the chiral fields admit a solution of the form (11).
The energy $E$ of the system can be written in the form
$$
\eqalign{
     E=\omega
        &+2\pi\int^\infty_0 r^2dr\big \{({d\sigma(r) \over dr})^2
         +({d\pi(r) \over dr})^2+{2 \over r^2}\pi^2(r)\big \}   \cr
         &+2\pi\lambda\int^\infty_0 r^2dr
                 \big[\sigma^2(r)+\pi^2(r)-v^2\big ]^2.
}
              \eqno (15)
$$
\par
A state will be bound if its energy $E$ is lower than the mass
$gv$ of one free fermion as $E/gv<1$;
a fermion orbital is bound and it decays exponentially at large distance if
its energy eigenvalue $\omega$ is such that $|\omega|<gv$.
In the strong coupling model, we would consider that the bound state
gives the mass for fermion.
The size and energy scale in such a way that once a bound state is found
with an approprite value of $g$, one can make the energy increase and
the size decrease proportionally with $v$[4].


\par
Before discussing a solution of the coupled set of equations (13),
it is instructive to consider a soluble model on chiral circle approximatiom.
When the constant $\lambda$ in the Lagrangian (8) is
large enough, the minimum energy occurs for chiral fields restricted to
the chiral circle $\sigma^2(r)+\pi^2(r)=v^2$.
In this situation it is possible to parametrize the chiral fields
by a chiral angle $\theta(r)$:
$$
    \sigma(r)=v\cos\theta(r), \qquad \pi(r)=v\sin\theta(r).  \eqno (16)
$$
When one fermion fills an orbital of energy
$\omega$, the energy (15) of the system can be expressed
in terms of the chiral angle as follow:
$$
    E=\omega+2\pi v^2\int^\infty_0 r^2dr \big\{({d\theta(r) \over dr})^2
                        +{2 \over r^2}\sin^2\theta(r)\big\}.   \eqno(17)
$$
\par
It is easy to establish that the equations for $\sigma(r)$ and $\pi(r)$
fields in (13) admits solutions which behave as $\theta(r)\to r-n\pi$
as $r\to 0$ and $\theta(r) \to 1/r^2$ as $r\to\infty$.
Let us consider a soluble model in which the chiral angle
$\theta(r)$ takes a 'trial' form $\theta(r)=-\pi(1-r/R)$ for $r<R$ and
$\theta(r)=0$ for $r>R$ carried out in Ref. 4 corresponding to the choice
$n=1$. The accuracy of this model is assessed in self-consistent
calculation, and it proves quite adequate for discussion of the physics.
\par
A numerical study of the Dirac equations in (13) with the linear form of
chiral angle has been carried out in Ref. 4. The energy eigenvalue $\omega$
is well described by the expression:
$$
    \omega={3.12 \over R}-0.94gv,  \eqno(18)
$$
valid in the range $2<R<12$.
By substitute the expression (18) for $\omega$ and the linear form
of the chiral angle into (17) we obtain a schematic expression for the
energy of the system as a function of $R$:
$$
     E={3.12 \over R}-0.94gv+2\pi v^2(1+{\pi^2 \over 3})R,
                      \eqno(19)
$$
This equation may be considered as an effective mass formula for the system.
\par
The equilibrium value of $R$ is the value which minimizes the energy (19).
One finds $R=0.34/v$, and the minimum energy is then given by
$$
     E_{\rm min}=v(18.33-0.94g).    \eqno(20)
$$
The coupling constant $g$ in this model must remain in the range
$9.5<g<19.5$ to make $E_{\rm min}<gv$ for the system bounded and
$E_{\rm min}>0$ for energy being non-negative.
\par
We see that, as the Yukawa coupling constant $f=\sqrt2 g$ increases,
the Higgs field around a fermion tends to go over from a uniform spatial
configuration to a hedgehog configuration which gives a decreasing mass
with increasing $f$.
We do think that the masses of lepton and quark are given by
the hedgehog configuration of Higgs field and not by the usual
uniform configuration (3) in the Weinberg-Salam model.
According to this idea, the energy of bound state is just the mass of
fermion as $E_{\rm min}=m_f$.
The hedgehog configurations of Higgs field give the masses of three
generations fermion by the following strongly Yukawa coupling constant
$y_i$ ($i$ for leptons and quarks) with
$$
       y_i={1 \over 0.94}(18.33\sqrt2 -f_i),  \eqno(21)
$$
where $f_i$ is the usual Yukawa coupling constant in (6).
The $y_i$ are given by:
$$
\eqalign{
    &y_{\nu_e}>27.572998, \quad y_e=27.572996,
     \quad y_u=27.572969, \quad y_d=27.572938,     \cr
    &y_{\nu_\mu}>27.572998, \quad y_\mu=27.572354,
     \quad y_c=27.565053, \quad y_s=27.571776,   \cr
    &y_{\nu_\tau}>27.572810, \quad y_\tau=27.562148,
     \quad y_t=26.498531, \quad y_b=27.546723.
}  \eqno(22)
$$
We found that, the quantitative differences of these Yukawa coupling
constants are behind the point except the top quark,
these Yukawa coupling constants are towards an unitary constant and
the hierarchy for Yukawa coupling constants is absent.
we should point out that all the values in (22) depend on the accuracy
of calculation, however, the results presented above do not qualitatively
depend on a particular choice for VEV $v$ and the calculating accuracy.


\par
In order to obtain more exact results for strongly Yukawa coupling constant,
we solve numerically the non-linear differential equations (13) for
eigenvalue $\omega$ under the boundary conditions (14) for bound state.
Integration of equation set (13), with normalization to one fermion,
was performed with the aid of the program COLSYS[6].
\par
In Fig. 1 we show the total energy of the system (15) as a function of
the Yukawa coupling constant for one bound fermion and for several
values of the Higgs mass.
We see that for a given Higgs mass, there exists a critical value of the
Yukawa coupling constant above which the fermion can form a stable bound
state (as a non-topological soliton[7]), which is lower in energy than
the normal mass of one free fermion. These results agree with those of [8].
For a given $f$, the energy of the bound state (soliton) increases
monotonically with the Higgs mass.
The energy of the bound state becomes nagative for the Yukawa coupling
constant $f\sim 21.0$ for the Higgs masses considered.
This value gives the up bound for the Yukawa coupling constant
which can be taken.
\par
As same as the soluble model, we think that the bound states give the masses
for lepton and quark.
{}From the (15) we see that the energy of the bound state (soliton) depends
on both Yukawa coupling constant $f$ and Higgs self-coupling constant
$\lambda$, So the mass of fermion also depends on both $f$ and $M_H$.
It is different from the standard picture of Weinberg-Salam model in which
the mass of fermion can be described alone by Yukawa coupling constant.
The strongly Yukawa coupling constants for lepton and quark are given
as the following where we take $M_H=v/2=0.174$TeV,
$$
\eqalign{
           &y_{\nu_e}>20.789611, \quad y_e=20.789611,
            \quad y_u=20.789595, \quad y_d=20.789577,   \cr
           &y_{\nu_\mu}>20.789611, \quad y_\mu=20.789236,
            \quad y_c=20.784979, \quad y_s=20.788911,   \cr
           &y_{\nu_\tau}>20.789502, \quad y_\tau=20.783310,
            \quad y_t=20.118602, \quad y_b=20.774090.
}     \eqno (23)
$$
These strongly Yukawa coupling constants tend to an unitary constant
as similar to the case in the soluble model but with smaller value.
Due to the numerical accuracy, the results represented above are not the
final values of Yukawa coupling constant for lepton and quark, however,
the strongly Yukawa coupling constants do not depend qualitatively on
the numerical accuracy.
\par
In Fig. 2 we show the radial functions for the bound state (soliton) with
Yukawa coupling constant $y_e$=20.789611 for electron mass in the case
of $M_H$=0.174TeV.
We see that the fermion is localized in a small region of the space,
while the Higgs fields approach its VEV, where $\sigma(r)\to -v$ and
$\pi(r)\to 0$, much slowly.
We can define the mean-square radius of the bound state as
$$
    r_0=\sqrt{<r^2>}, \qquad  <r^2>=\int_0^\infty r^4dr[G^2(r)+F^2(r)].
             \eqno (24)
$$
The typical radius of the bound state for the lepton and quark is
$r_0\sim 0.67$/TeV with $M_H$=0.174TeV.
This size is fitting for the limit value of recent experimental data[9].
The typical intrinsic energy scale in this strong coupling model is
$M_0=yv/\sqrt2\sim3.6$TeV.
This leads us to the interesting possibility of probing the size of
lepton and quark in next generation collider.
\par
As shown in Fig. 2, the hedgehog configuration of Higgs fields
[$\sigma(r)$, $\pi(r)$] obviously appears in the region of small distance
(smaller than 1.0/TeV) and tends to the uniform configuration in large
distance (much bigger than 1.0/TeV).
On the other hand, the electroweak gauge symmetry is broken at the energy
scale of about 100GeV (in the distance of 10.0/TeV), in which the hegdehog
configuration of Higgs field is close to the normal vacuum
configuration (3).
So the strong coupling model produces the masses of $W$ and $Z$ almost
as same as Weinberg-Salam model.
However, there is a little different from the case of Weinberg-Salam model
due to the fact that field $\pi(r)$ is not exactly zero in the region of
one hundred GeV. Then, it is possible to research this little difference in
recent experiments.
\par
The above physics remains roughly unaltered when we turn on a weakly coupled
gauge sector[3].
It is shown that for small gauge coupling the bound state fermion continues
to be a good approximation for the one fermion ground state in the standard
model.
Nolte and Kunz[8] have taken a numerical calculation and show that the change
in energy for the strong Higgs-fermion bound state due to the presence of the
gauge field is negligible.


\par
In conclusion, we have attempted to give the masses of lepton and quark
associated with the electroweak standard model when it carries a strongly
coupled Higgs-fermion sector.
Within the classical approximation, we have shown that the strongly Yukawa
coupling constants for lepton and quark are towards to an unitary constant
and the hierarchy for Yukawa coupling constants is absent.
On the other hand, in the strong coupling model, the masses of $W$ and $Z$
produced by the hedgehog configuration of Higgs field keep almost no change.
The mass of fermions depends slightly on the mass of Higgs boson in the tree
level, then, this relation should shed light on hunting for the Higgs boson.
\par
The strong coupled Higgs-fermion bound state (soliton) may be unstable with
the quantum radiative corrections[10].
This possible quantum instability should be the common problem in the
non-point model for lepton and quark.
It is due to the fact that the resulting bound states have the sizes,
which are so much smaller than their Compton wavelengths.
This fact is only understandable dynamically if there exists an approximate
symmetry which forces certain bound states to have nearly zero mass.
On the other hand, the strongly Yukawa coupling in the standard model will
violate the vacuum stability[5] due to the radiative corrections, then,
we should improve the standard model with new physics in the TeV region.
So, in order to stabilize the bound states for fermion and Higgs vacuum
structure against quantum effects, it should introduce new symmetry into
the standard model, such as supersymmetry.


\par
The author is indebted to Dr. D. H. Zhang for help in drafting on computer.

\vfill\eject

\centerline{\bf REFERENCES }
\vskip 1pc
\item{1} S. Glashow, \np {\bf 22}, 579(1961);
\item{ } S. Weinberg, \prl {\bf 19}, 1264(1967);
\item{ } A. Salam, in: Elementary Particle Theory, ed. N. Svartholm,
                                                            367(1968).
\item{2} CDF Collaboration, F. Abe $et\quad al.$, \prl {\bf 74}, 2626(1995).
\item{3} R. Johnson and J. Schechter, \pr {\bf D36}, 1484(1987);
\item{ } V. Soni, B. Moussallam and S. Hadjitheodoridis, \pr {\bf D39},
                     915(1989).
\item{4} S. Kahana, G. Ripka and V. Soni, \np {\bf A415}, 351(1984);
\item{ } M. Birse and M. Banerjee, \pr {\bf D31}, 118(1985).
\item{5} P. Huang, \prl {\bf 42}, 873(1979);
\item{ } H. Politzer and S. Wolfram, \pl {\bf 82B}, 442(1979);
                                         {\bf 83B}, 42(1979).
\item{6} U. Ascher, J. Christiansen and R. Russell, Math. Comput.
                                              {\bf 33}, 659(1979).
\item{7} R. Friedberg and T. Lee, \pr {\bf D15}, 1694(1977);
                                      {\bf D16}, 1096(1977).
\item{8} G. Nolte and J. Kunz, \pr {\bf D48}, 5905(1993);
\item{ } G. Petriashvili, \np {\bf B378}, 468(1992).
\item{9} Partical Data Group, \pr {\bf D50}, 1227(1994).
\item{10} G. Anderson, L. Hall and S. Hsu, \pl {\bf 249B}, 505(1990);
\item{ } J. Bagger and S. Naculich, \prl {\bf 67}, 2252(1991).


\vskip 2pc
\centerline{\bf FIGURE CAPTIONS}
\vskip 1pc

\item{} ${\bf FIG. 1}$ The bound state energy (fermion mass) is shown
        as a function of the Yukawa coupling constant for Higgs mass
        $M_H=0.174$TeV (solid), $M_H=0.522$TeV (dot-dashed) and
        $M_H=0.870$TeV (dashed) for one bound fermion. The dotted
        line presents the usual Yukawa coupling case.
\vskip 1pc
\item{} ${\bf FIG. 2}$ The fermion field functions $G(r)$ (solid) and $F(r)$
        (dotted) and Higgs field functions $\sigma(r)$ (dot-dashed)
        and $\pi(r)$ (dashed) of the bound state (soliton)
        for electron are shown with respect to the distance $r$
        for $M_H=0.174$TeV and Yukawa coupling constant $f=20.789611$.

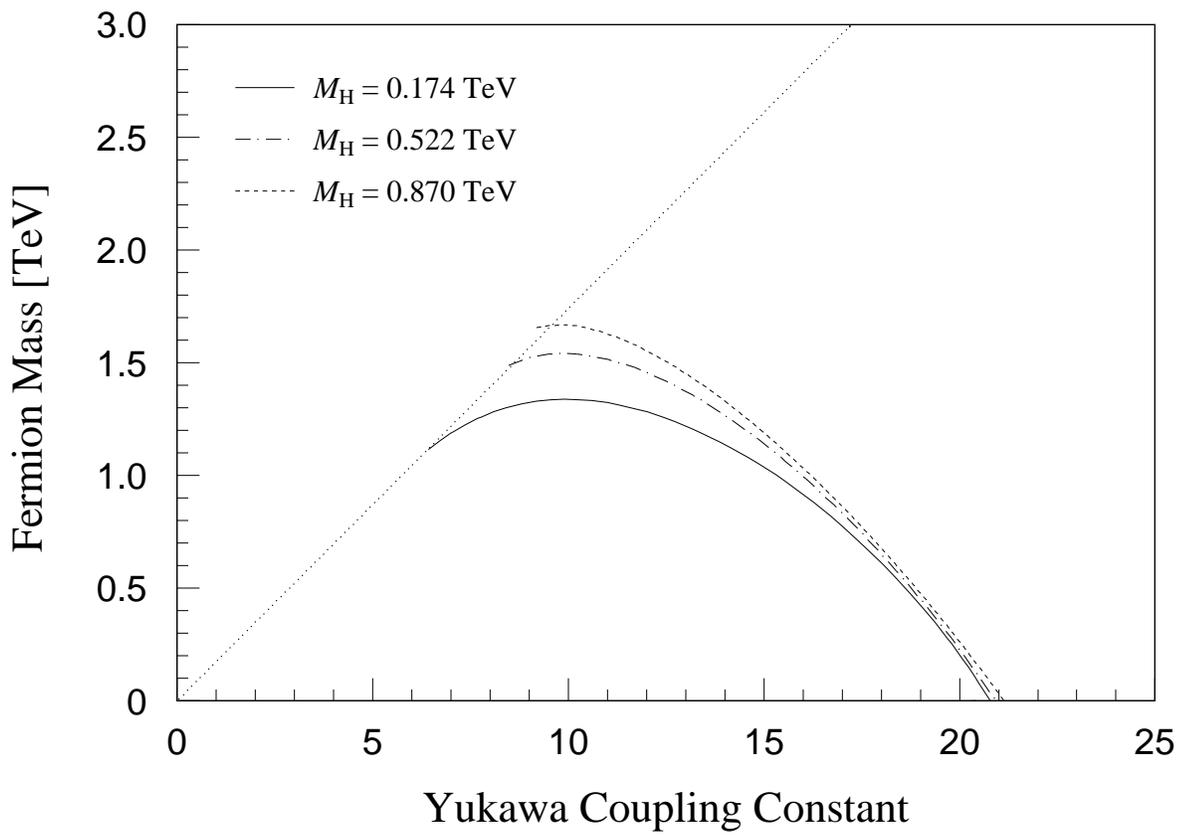
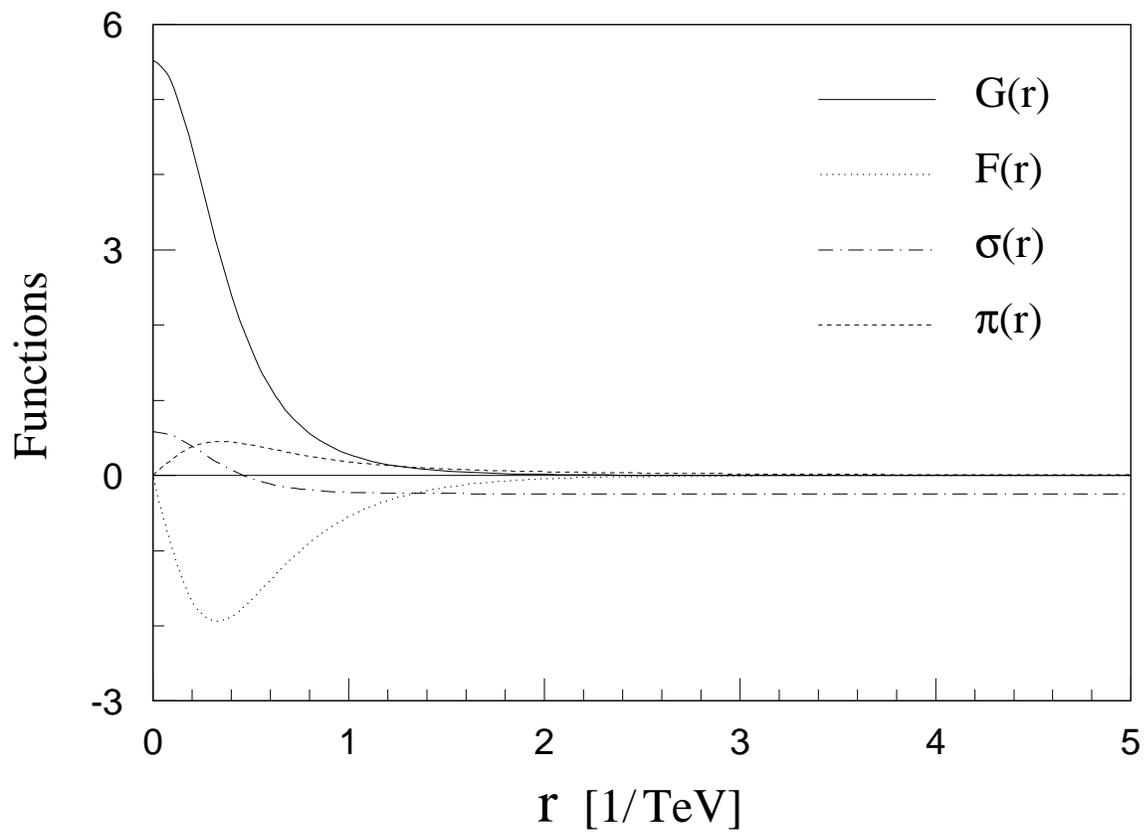
\end